\documentclass[preprint]{aastex61}

\newcommand\aastex{AAS\TeX}

\submitjournal{ApJL}

\shorttitle{\aastex\ Long-term dynamical evolution of MPs}
\shortauthors{Dalessandro et al.}

\begin{document}

\title{A family picture: tracing the dynamical path of the structural properties of multiple populations in globular clusters}

\correspondingauthor{Emanuele Dalessandro}
\email{emanuele.dalessandro@inaf.it}


\author{Emanuele Dalessandro}
\affiliation{INAF - Astrophysics and Space Science Observatory Bologna, Via Gobetti 93/3 40129 Bologna - Italy}

\author{M. Cadelano}
\affiliation{Dipartimento di Fisica e Astronomia,  Via Gobetti 93/2 40129 Bologna - Italy}
\affiliation{INAF - Astrophysics and Space Science Observatory Bologna, Via Gobetti 93/3 40129 Bologna - Italy}

\author{E. Vesperini}
\affiliation{Department of Astronomy, Indiana University, Swain West, 727 E. 3rd Street, IN 47405 Bloomington - USA}

\author{S. Martocchia}
\affiliation{European Southern Observatory, Karl-Schwarzschild-Stra$\beta$e 2, D-85748 Garchin bei Munchen, Germany}
\affiliation{Astrophysics Research Institute, Liverpool John Moores University, 146 Brownlow Hill, Liverpool L3 5RF, UK}

\author{F. R. Ferraro}
\affiliation{Dipartimento di Fisica e Astronomia,  Via Gobetti 93/2 40129 Bologna - Italy}
\affiliation{INAF - Astrophysics and Space Science Observatory Bologna, Via Gobetti 93/3 40129 Bologna - Italy}

\author{B. Lanzoni}
\affiliation{Dipartimento di Fisica e Astronomia,  Via Gobetti 93/2 40129 Bologna - Italy}
\affiliation{INAF - Astrophysics and Space Science Observatory Bologna, Via Gobetti 93/3 40129 Bologna - Italy}

\author{N. Bastian}
\affiliation{Astrophysics Research Institute, Liverpool John Moores University, 146 Brownlow Hill, Liverpool L3 5RF, UK}


\author{J. Hong}
\affiliation{Department of Astronomy, Indiana University, Swain West, 727 E. 3rd Street, IN 47405 Bloomington - USA}
\affiliation{Kavli Institute for Astronomy and Astrophysics, Peking University, Yi He Yuan Lu 5, HaiDian District, Beijing 100871, China}

\author{N. Sanna}
\affiliation{INAF - Osservatorio Astrofisico di Arcetri, Largo Enrico Fermi 5, 50125 Firenze, Italy}






\begin{abstract}
We studied the spatial distributions of multiple stellar populations
(MPs) in a sample of 20 globular clusters (GCs) spanning a broad range
of dynamical ages. The differences between first-population (FP) and
second-population (SP) stars were measured by means of the parameter
$A^+$, defined as the area enclosed between their cumulative radial
distributions.  We provide the first purely observational evidence of
the dynamical path followed by MPs from initial conditions toward a
complete FP-SP spatial mixing.   
Less dynamically evolved clusters have SP stars more centrally concentrated than FPs, 
while in more dynamically evolved systems the spatial differences between FP and SP stars decrease 
and eventually disappear.
By means of an appropriate comparison with a
set of numerical simulations, we show that these observational results
are consistent with the evolutionary sequence expected by the
long-term dynamical evolution of clusters forming with an
initially more centrally concentrated SP sub-system.  This result is
further supported by the evidence of a trend between $A^+$ and the
stage of GC dynamical evolution inferred by the ratio between the
present-day and the initial mass of the cluster.

\end{abstract}




\section{Introduction} \label{sec:intro}

The presence of sub-populations differing in terms of their light-element abundances (e.g. He, C, N, O, Na, Mg, Al) while having the same iron (and iron-peak) content (hereafter multiple stellar populations - MPs) is a key general property of globular clusters (GCs; see \citealt{bastian18} for a recent review). 
In fact, MPs are observed in nearly all old ($t>2$ Gyr) and relatively massive systems ($M>10^4 M_{\odot}$), both in the Milky Way and in external galaxies (e.g., \citealt{mucciarelli08,larsen14,dalessandro16}).

MPs are characterized by specific light-element chemical abundance patterns like C-N, Na-O, Mg-Al anti-correlations.
Stars sharing the same chemical abundances as the surrounding field stars (Na-poor/O-rich, CN-weak) are commonly classified as first-population (FP), while Na-rich/O-poor, CN-strong stars are referred to as second-population (SP). 
Light-element chemical abundance variations can have an impact on both  the stellar structure and atmosphere thus 
producing a variety of features (such as broadening or splitting of different evolutionary sequences) in color-magnitude diagrams (CMDs) when appropriate optical and near-UV bands are used \citep{sbordone11,piotto15,milone17}. 
It has been shown that the fraction of SP stars and the amplitude of the light-element anti-correlations depends on the present-day cluster mass (e.g. \citealt{carretta10,schiavon13,milone17}), with relatively small systems ($M<10^5 M_{\odot}$) typically having a fraction of $\sim 40\%-50\%$ of SP stars, which then increases to $\sim 90\%$ for the most massive ones. Light-element inhomogeneities appear to decrease also as a function of cluster age, becoming undetectable for cluster younger than $\sim 2$ Gyr (\citealt{martocchia18a}), 
although the exact role of age is currently not clear yet.

MPs are believed to form during the very early epochs of GC formation and evolution ($\sim 10-100$ Myr, but see \citealt{martocchia18b} for recent observational constraints on this aspect). A number of scenarios have been proposed over the years to explain their formation, 
however their origin is still strongly debated \citep{decressin07,dercole08,bastian13,denissenkov14,gieles18,calura19}. 

The kinematical and structural properties of MPs can provide key
insights into the early epochs of GC evolution and formation.
In fact, one of the predictions of MP formation models (see
e.g. \citealt{dercole08}) is that SP stars form a centrally segregated
stellar sub-system possibly characterized by a more rapid internal
rotation \citep{bekki11} than the more spatially extended FP
system.
Although the original structural and kinematical differences between FP and SP stars are gradually erased 
during  GC long-term dynamical evolution (see e.g. \citealt{vesperini13,henault15,miholics15,tiongco19}), some clusters are expected to still retain some memory of these initial differences in their present-day properties.

Indeed, sparse and inhomogeneous observations show that MPs are characterized by quite remarkable differences in their relative structural parameters/radial distributions \citep{lardo11,dalessandro16,massari16,simioni16}, different degrees of orbital anisotropy \citep{richer13,bellini15}, 
different rotation amplitudes \citep{cordero17} and significantly different binary fractions \citep{lucatello15,dalessandro18a}. 
However, so far the lack of a homogeneous and self-consistent  study of MP kinematical and structural properties for a statistically representative sample of clusters has hampered our ability to build an observational picture to test and constrain models for the formation and evolutionary history of GCs.

In this Letter we use the $A^+$ parameter
(originally introduced for blue straggler star studies; \citealt{alessandrini16,lanzoni16})
to quantify the
  differences in the radial distributions of FP and SP stars for a
  large sample of GCs in different stages of their dynamical
  evolution measured here by the ratio $N_h = t/t_{rh}$ between 
 the cluster age $t$ and its current 
half-mass relaxation times ($t_{rh}$). 
A comparison of our results with those of numerical simulations
following the dynamical evolution and spatial mixing of MPs allows us
to draw, for the first time, an observational picture of the
evolutionary path of FP and SP structural properties.

\section{Sample definition and population selection}

For the present analysis we mainly used the publicly available photometric catalogs of Galactic GCs presented 
in \citet[see also \citealt{piotto15}]{nardiello18} and observed through proposals GO-13297, GO-12605 and GO-12311
(PI: Piotto) with the HST WFC3/UVIS camera in the F275W, F336W and F438W bands and with the HST ACS/WFC 
under proposal GO-10775 (PI: Sarajedini) in the F606W and F814W filters.  
We limited our analysis only to systems for which the available HST catalogs cover at least 2 cluster half-light radii ($r_h$) allowing us to probe a region large enough to capture possible differences between the SP and FP spatial distributions.

With the adopted selection we are able to include in our sample
  15 GCs, most of which have $N_h>7-8$. To further extend our analysis and include 
clusters with smaller values of $N_h$, which is essential for the goals of our study, we complemented our 
data-set with the wide-field photometric catalog (that includes U, B, V and I bands)
  published by \citet{stetson19} for the low-mass cluster NGC288, the
  Stromgren photometry of NGC5272 (M3) presented by \citet{massari16}
  and the combined HST and ground-based wide-field catalog of NGC6362
  published in \citet{dalessandro14}.
Finally, we included also two
extra-galactic systems, namely NGC121 in the Small Magellanic Cloud
and NGC1978 in the Large Magellanic Cloud.  The HST photometry of
these two clusters was presented in \citet{dalessandro16} and
\citet{martocchia18a} respectively.
It is important to stress that to make the MP separation and selection
as straightforward/clear as possible, only clusters with
intermediate-high metallicity\footnote{It is well known that the
  amplitude of color variations caused by the effect of light-element
  anti-correlations decreases with metallicity. Thus, photometric broadenings or
    splittings of the evolutionary sequences in the CMD are harder to
    detect in metal-poor systems.}, low reddening, relatively low
field contamination and with a well populated red giant branch were
added to the initial list of 15 GCs.  
With such a combination our sample counts 20 GCs covering (see Table~1) 
a wide range in metallicity ($-0.4<[Fe/H]<-2$) and present-day mass ($3.6\times 10^4 M_{\odot}<M<1.4\times10^6 M_{\odot}$), 
which are well representative of the population of Galactic and Magellanic Cloud GCs, with the exception of the lower mass systems in the Clouds.  More importantly to the present analysis, the sample covers the full range of dynamical stages 
derived for Galactic and Magellanic Cloud clusters ($1<N_h<80$).

\begin{figure}
\plotone{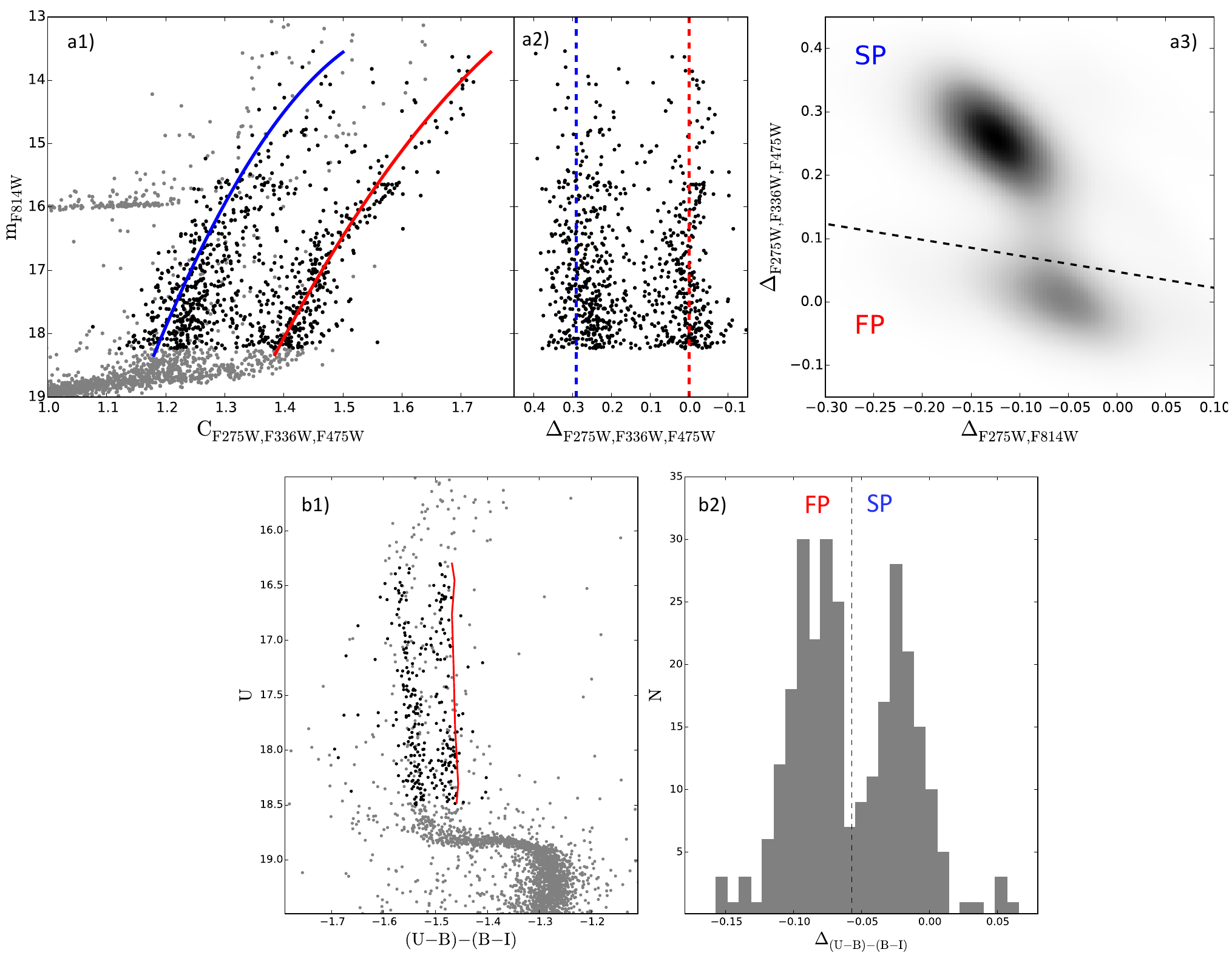}
\caption{Panel a1): ($m_{F814W}, C_{F275W,F336W,F438W}$) CMD of NGC6541. Data are from \citet{nardiello18}. 
The red and blue lines represent the two fiducial lines at the edge of the RGB. Black dots are stars selected as described in Section~2. a2): verticalized $m_{F814W}, \Delta_{F275W,F336W,F428W}$ distribution of RGB stars with respect to the fiducial lines. a3): the derived 
($\Delta_{F275W,F814W}, \Delta_{F275W,F336W,F438W}$) diagram. The black dashed line marks the boundary between FP and SP stars. 
Panel b1): ($U, (U-B)-(B-I)$) CMD of NGC288. Data are from \citet{stetson19}. The red line represents the fiducial line at the bluer 
edge of the RGB. b2): distribution of the verticalized color $\Delta_{(U-B)-(B-I)}$. As before the black dashed line marks the limit adopted to separate FP from SP stars.}
\end{figure}

For the clusters for which we used the photometric catalogs published by \citet{nardiello18}, 
MPs were selected along the red giant branch (RGB) in the ($\Delta_{F275W,F814W}, \Delta_{F275W,F336W,F438W}$) diagram, 
the so called ``chromosome map'', following the same approach used by \citet{milone17} and schematically shown in Figure~1 
(panels a). Briefly, we verticalized the distribution of RGB stars in the  ($m_{F814W}, C_{F275W,F336W,F438W}$) (where $C_{F275W,F336W,F438W}=(m_{F275W}-m_{F336W}) - (m_{F336W}m_{F438W})$) and 
($m_{F814W}, m_{F275W}-m_{F814W}$) diagrams with respect to two fiducial lines at the blue and red edges of the RGB 
in both CMDs (Figure~1 panels a1 and a2). The combination of the 
two verticalized distributions ($\Delta_{F275W,F814W}$ and $\Delta_{F275W,F336W,F438W}$) gives 
the ``chromosome map" (Figure 1 panel a3).
Only stars with a membership probability $>75\%$ and with 
quality flags $>0.9$ in all bands were used (see \citealt{nardiello18} for details).

For NGC121, NGC6362, M3 and NGC1978 we adopted the same sub-population selections described in \citet{dalessandro14,dalessandro16,massari16,martocchia18a} respectively.

For the case of NGC288, we used a two-step approach. For stars at a cluster-centric distance $R<100\arcsec$ we used the HST catalog published by \citet{nardiello18} and the selection criteria described before. 
For the external region we first matched the ground-based catalog with Gaia DR2 data.
Cluster bona-fide stars were selected based on their Gaia proper motions. We assumed ($\mu_{\alpha}=4.24, \mu_{\delta}=5.65)$ mas/yr as cluster mean motion \citep{helmi18} and we selected stars at distance $d<1.5$ mas yr$^{-1}$ in the vector-point diagram. 
RGB likely cluster members
were verticalized in the ($U, (U-B)-(B-I)$) CMD with respect to a fiducial line on the blue edge of the RGB (Figure~1 panel b1; see also \citealt{monelli13}). The resulting distribution is clearly bimodal (panel b2). Stars redder/bluer than $\Delta_{(U-B)-(B-I)}=-0.55$ were selected as FP/SP stars.  

It is important to note that, while in general, the adoption of different filter combinations for FP and SP classifications can introduce some bias, this is not the case for the specific targets in our sample for which both ground-based photometry and the HST ``chromosome-map'' are available,
namely NGC288, NGC6362 and M3.  In fact, 
we have verified, by using the stars in common between the available HST and wide-field catalogs, that there is a nice match between 
the two sub-population selections thus ensuring homogeneity of the different samples.

\begin{figure}
\plotone{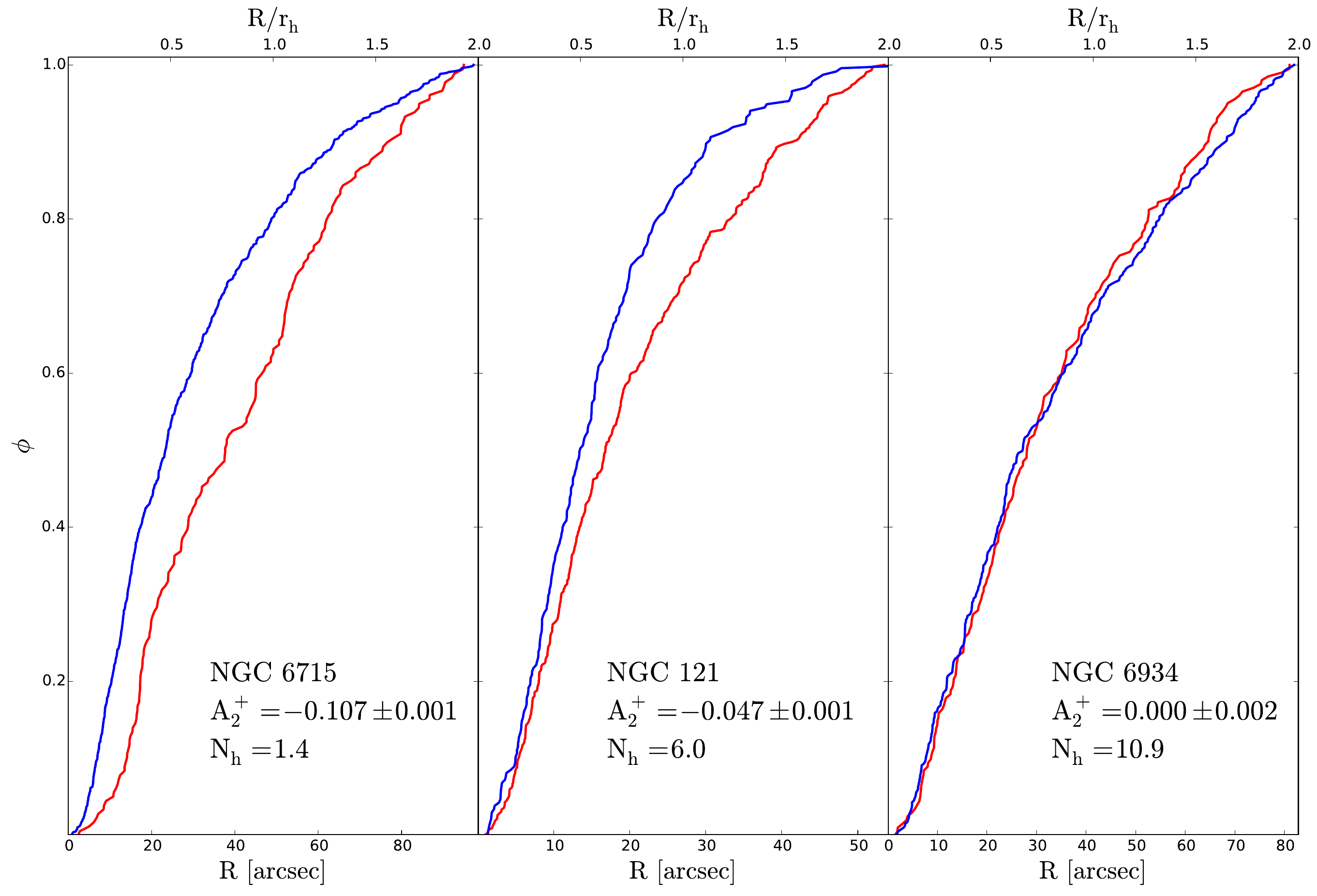}
\caption{Cumulative radial distributions of FP (red) and SP (blue) for three
  representative clusters: M54 is
    one of the clusters with the smallest value of $N_h$ in the
    sample, and it shows a very negative value of $A^+_2$,
  while NGC 6934 is an example of fully radial mixed cluster and
   NGC 121 is an intermediate case.}
\end{figure}

\section{Radial distribution of multiple populations and empirical derivation of the parameter $A^+$}  

We derived the cumulative radial distributions of the selected sub-populations by using the cluster centers 
reported in \citet{ferraro12} and \citet{lanzoni16} and references therein for the clusters in common, and those listed in \citet{goldsbury10} for the other Galactic GCs. For NGC121 and NGC1978 we used the centers derived by \citet{dalessandro16} and \citet{martocchia18a} respectively. 

In order to obtain a homogeneous measure of the differences between the SP and FP spatial distributions 
we have used the $A^+$ parameter introduced by \citet{alessandrini16} and \citet{lanzoni16} in the context of the study the spatial segregation of blue straggler stars.
In our study $A^+$ is calculated as the area enclosed between the cumulative radial
  distributions of FP and SP stars, $\phi_{FP}(R)$ and $\phi_{SP}(R)$,
  respectively:
\begin{equation}
A^+(R)=\int_{R_{min}}^R (\phi_{FP}(R^{'})-\phi_{SP}(R^{'}))dR^{'}
\end{equation}

\begin{figure}
\plotone{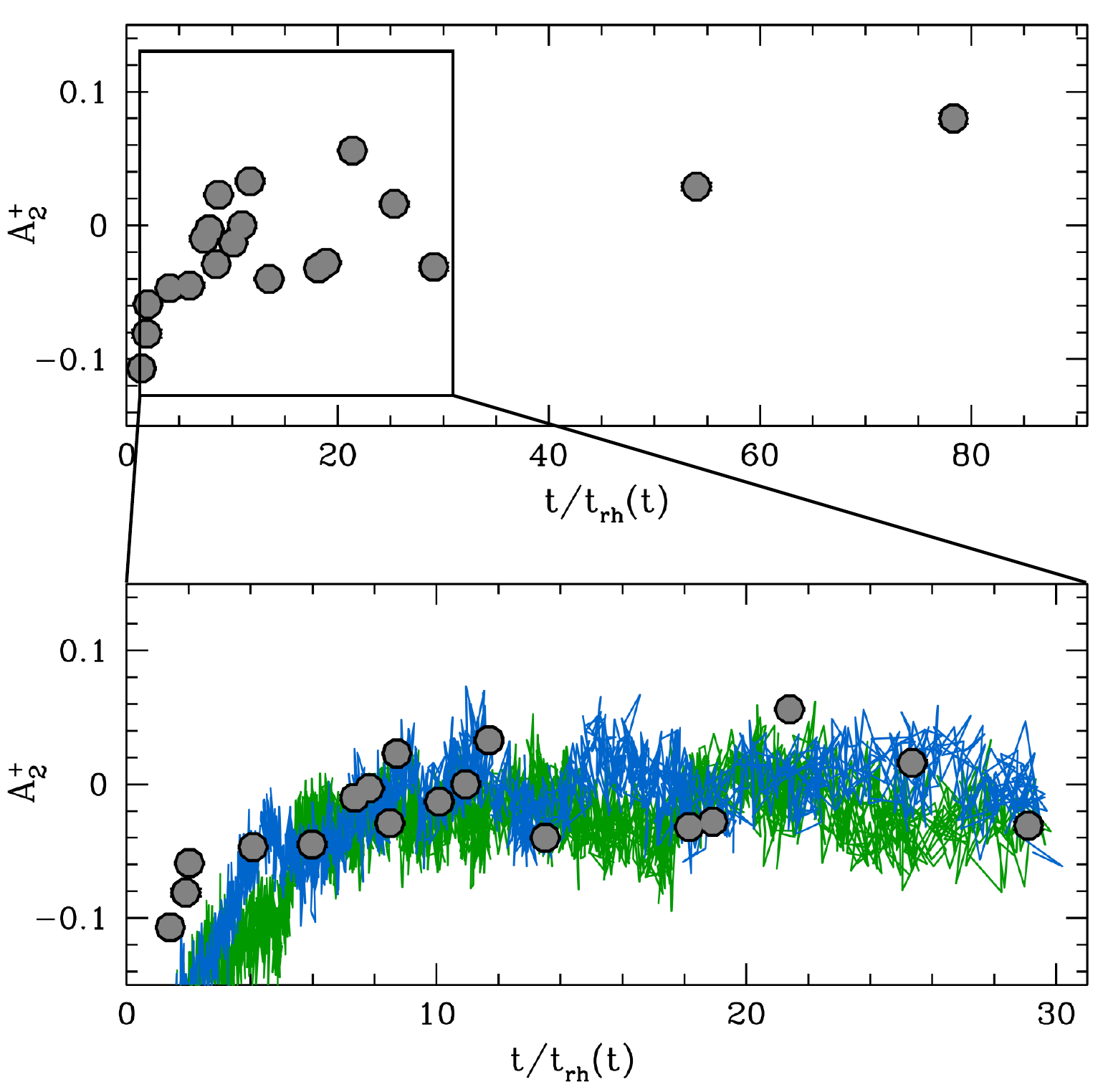}
\caption{Upper panel: distribution of $A^+_2$ as a function of $t/t_{rh}$ ($N_h$) for all the clusters in the sample.
Bottom panel: zoom on the distribution of cluster with $N_h<30$. 
Results from $N$-body models are overplotted to the observations. Blue and green curves represent models starting with 
a SP 5 and 10 times more centrally concentrated than FP respectively. }
\end{figure} 

where R is the distance from the cluster center.
With such a definition, a more centrally concentrated SP yields
negative values of $A^+$.  By construction $A^+$ depends on the
considered cluster-centric distance and therefore a meaningful
cluster-to-cluster comparison requires that the parameter is measured
over equivalent radial portions in every system.
As shown in numerical studies (see e.g. \citealt{vesperini13}), spatial mixing is achieved first 
in a cluster's inner regions
and later in the cluster's outskirts. Therefore capturing a complete dynamical picture of the mixing
process in a given cluster would require a wide radial coverage possibly
extending to the cluster's outermost regions, which retain memory of the initial spatial differences for a longer time.
With this in mind, we decided to measure $A^+$ within 2
  $r_h$ from the cluster center ($A^+_{2}$).
This limit represents a compromise between radial coverage and cluster sample size.
We adopted the values
of $r_h$ reported by (\citealt{harris96} - 2010 version) for all the
Galactic clusters, while we used \citet{glatt11} for NGC121. For
NGC1978 we derived $r_h=31.5\arcsec$ by fitting its number count
density profile (derived by using the HST catalog) with a single-mass
\citet{king66} model.

Uncertainties on the derived values of $A^+$ have been obtained by applying a jackknife bootstrapping technique \citep{lupton93}.  The results are reported in Table~1.

\begin{figure}
\plotone{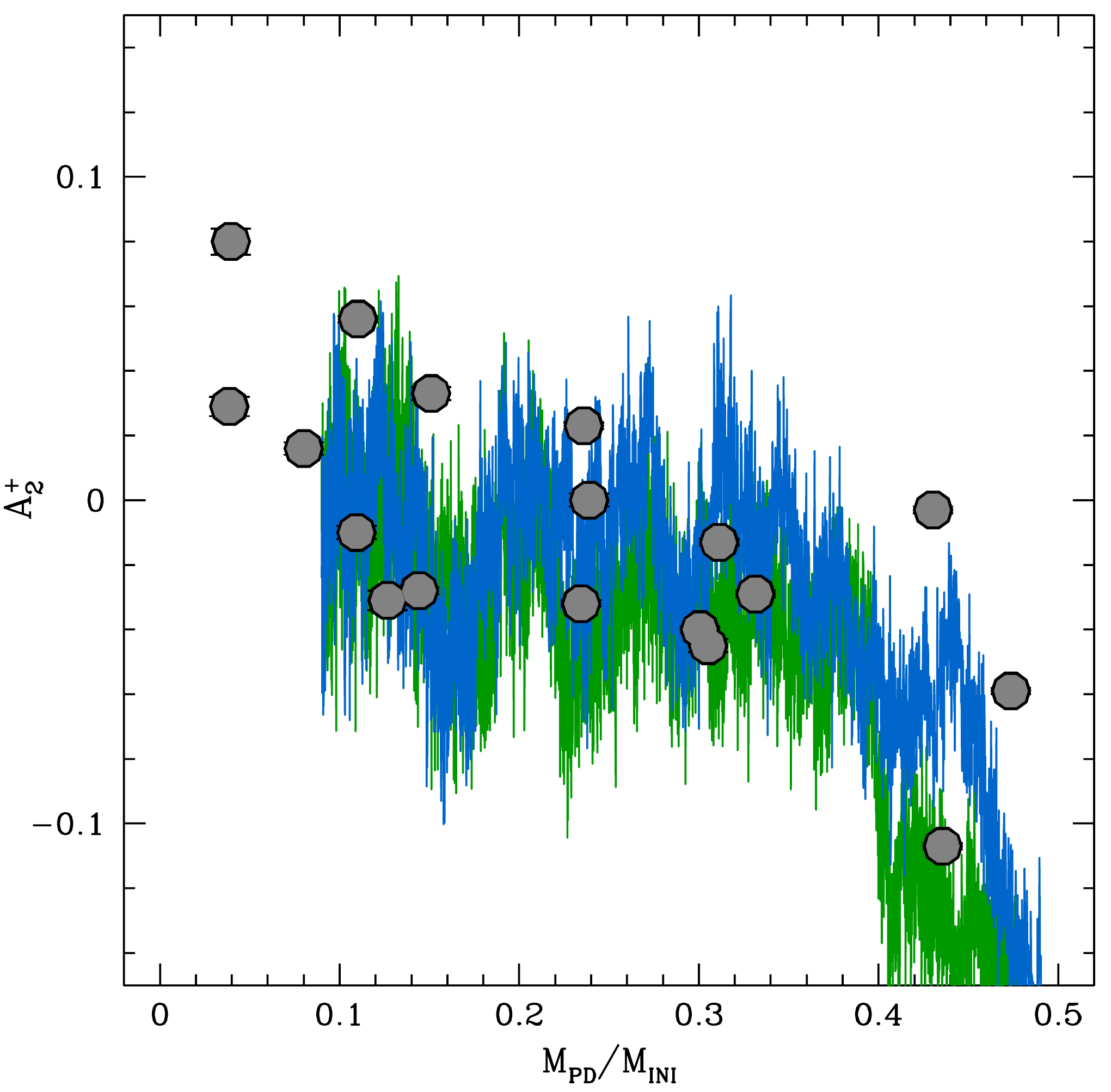}
\caption{Distribution of $A^+_2$ as a function of the ratio
    between the present-day and the initial cluster mass ($M_{\rm
    PD}/M_{\rm ini}$) obtained by \citet{baumgardt19}. Blue and green
  curves represent the same models shown in Figure 3.}
\end{figure} 

\section{Results}

The MP radial distributions in the targeted clusters appear to be quite different from one case to the other.
However, in general we can identify two main behaviors: in about half of the sample, SP stars are more centrally concentrated than FPs,
in the other clusters there is no significant difference between the FP and SP distributions. 
As a result, 
the derived values of  $A^+_{2}$ cover a quite large range, from a minimum of $\sim-0.107\pm 0.006$ for NGC6715 (M54) to $\sim0.080\pm0.016$ for NGC6717  (Table~1). The cumulative radial distributions for three systems with different behaviors are shown in Figure~2 as an example.

For every cluster we determined $N_h$ by adopting the ages
  derived by \citet{dotter10} for Galactic GCs and by
  \citet{martocchia18a} and \citet{glatt11} for NGC1978 and NGC121
  respectively, while the values of $t_{rh}$ are taken from
  \citet{harris96} and \citet{glatt11} for NGC121.
For NGC1978 we derived Log$(t_{rh})=9.02$ (where $t_{rh}$ has been calculated as in \citealt{harris96}).
Figure 3 shows the distribution of $A^+_2$ as a function of
  $N_h$. The $A^+_2$ parameter increases almost linearly up to $N_h
  \sim 10$, reaching values close to 0 where FP and SP stars are
  (almost) fully radially mixed, then it shows an almost constant
  distribution for older dynamical ages up to $N_h\sim80$.

The general trend shown in Figure 3 suggests that SP stars are
  significantly more concentrated than FPs in systems with $N_h<
  8-10$, while MP radial distributions do not show significant
  differences for clusters in more advanced stages of their 
  dynamical evolution (with $N_h>10$). The only two exceptions are NGC
  6093 (M80) and NGC 6717, which are the systems in the sample
  characterized by most positive values of $A^+$.
The MP radial distribution of M80 has been analyzed in detail and 
extensively discussed in \citet{dalessandro18b}. 

To illustrate the expected evolution of $A^+_2$ as a function of
  $N_h$, in Figure 3 (bottom panel) we show the time evolution of
$A^+_2$ obtained from $N$-body simulations following the long-term
dynamical evolution of two MP clusters in which the SP is initially 5
and 10 times more centrally concentrated than the FP one.  
The simulations start with 50000 stars equally split between FP and SP and follow a cluster internal 
evolution and mass loss due to the combined effects of two-body relaxation and tidal truncation. The simulations have been presented in \citet{vesperini18} and \citet{dalessandro18b} and we refer to those papers for further details.
Here we use these simulations to
explore the role of internal two-body relaxation and the interaction
of the external tidal field of the host galaxy in the evolution of
$A^+_2$ as a function of $N_h$. 
We point out that the simulations presented here are still idealized and not meant to model any specific 
cluster in detail, but they serve to illustrate the general evolutionary trend expected for $A^+_2$
as the SP and the FP mix. 
Detailed models aimed at reproducing the properties of specific clusters would require more realistic simulations.

Since the $N$-body models start with a more centrally concentrated SP
radial distribution, the simulations have initially negative values of
$A^+_2$. As the FP and SP stellar sub-systems evolve (i.e. $N_h$
increases) the two populations gradually mix and, as a consequence,
$A^+_2$ increases evolving toward  zero, which represents the value corresponding
to a fully radially mixed configuration.  Although the simulations are
still simplified,
they follow the general $A^+_2$ trends. This suggests that the different shapes of MP radial 
distributions and the trend found in this study are the result of the effects of the 
long-term dynamical evolution in clusters formed with an initially more centrally concentrated SP stellar sub-system. 

It is important to note that in this comparison FP and SP are assumed to have the same He abundance or only small mean variations ($\Delta Y<0.01-0.02$).
Indeed, this is observed to be the case in the vast majority of GCs (see for example \citealt{dalessandro13})
with only a few exceptions in our sample, such as NGC2808 \citep{piotto07}, M80 \citep{dalessandro18b}, NGC7078 (M15) and M54 \citep{milone18}.

In Figure 4 we show the dependence of $A^+_2$ on the ratio between the present-day and the initial
  cluster mass ($M_{\rm PD}/M_{\rm ini}$), as estimated by
  \citet{baumgardt19}.  Although it is important to emphasize that
much caution should be used in taking $M_{\rm PD}/M_{\rm ini}$ ratios
at face value because of the underlying strong assumptions made to
derive them, and the possible missing contribution of related
effects\footnote{Examples of the missing contribution are early
    time-variation of the external potential or other mechanisms
    related to a cluster's response to early evolutionary processes
    (e.g. gas expulsion, mass loss due to stellar evolution,
    interactions with giant molecular clouds)}, they nevertheless
  provide a measure of the evolutionary stage of a cluster and its
  degree of mass loss due to two-body relaxation and the interaction
  with the Galactic potential.  Our data show a significant
  correlation (Spearman's rank correlation coefficient $r\sim-3.7$)
  between $A^+_2$ and $M_{\rm PD}/M_{\rm ini}$: clusters with small
  value of $M_{\rm PD}/M_{\rm ini}$ (i.e. systems that lost a larger
  fraction of their original mass) tend to have their MPs
    spatially mixed.  Interestingly, such a behavior is also
  reproduced (at least qualitatively) by our $N$-body models, thus
  demonstrating that the fraction of mass lost is a key ingredient of
  the MP spatial mixing process (see the discussion on this
    issue in \citealt{vesperini13,henault15,miholics15}).

Not surprisingly (because of the known dependence with the dynamical parameters used before) 
we find that $A^+_2$ nicely anti-correlates with
the present-day mass ($M_{PD}$ from \citealt{baumgardt18}).

\section{Conclusions}

The variations of the MP radial distributions
as a function of the evolutionary stage in the clusters' dynamical evolution shown in this Letter provides {\it
  the first observational evidence of the dynamical path followed by
  MPs from their initial conditions toward a complete spatial mixing.}

Our study has revealed a clear trend of the difference between the SP and FP spatial radial distributions ($A^+_2$) 
and globular cluster dynamical evolution, as constrained 
by both the ratio of a cluster's age to its half-mass relaxation timescale and the ratio of a cluster's present-day to its initial mass. This is the first time that observational constraints on the evolutionary path of the MP structural differences are set and put in the framework of star cluster dynamical evolution.

Although additional work is needed to constrain in detail the initial physical properties of MPs both observationally and in the context of different theoretical formation models, our results provide a global view of the evolution of the MP structural properties. They lend support to an interpretation of the different degrees of spatial mixing observed in various clusters in terms of dynamical evolution of systems in which the SP formed more centrally concentrated than the FP. At the same time, the empirical evolutionary sequence found in our analysis also provides a key constraint for models exploring the long-term dynamics of MPs, which is an important aspect of the study of MP clusters.

The result presented here has important implications also for the interpretation of other kinematical features observed in MPs, such as their rotation patterns and anisotropy profiles, and therefore is key to shed light on the physical initial conditions that brought to the formation of MPs.

An extension of the present analysis, mainly including a
  larger sample of less dynamically evolved clusters, is needed to
  further confirm and sharpen the picture emerging from our study.

In addition, a systematic combination of structural and kinematic information of MPs is an essential step to properly interpreting observational data, as well as testing the key elements of theoretical scenarios of cluster formation and evolution.

\startlongtable
\begin{deluxetable}{ccccccccc}
\tablecaption{GC $A^+$ info \label{tab:table}}
\tablehead{
\colhead{Cluster}  & \colhead{$A^{+}_{2}$} & \colhead{$\epsilon$}  & \colhead{$\log(t)$} & \colhead{$\log(t_{rh})$} & \colhead{$M_{PD} \, (\times 10^5 M_{\odot})$} & \colhead{$r_h \, (\arcsec)$} & \colhead{$R_{GC} \, (kpc)$} & \colhead{$[Fe/H]$} }
\startdata
NGC121   &  -0.047 & 0.001 & 10.021 & 9.53 &   3.42 & 27.0  & 61.9  & -1.28  \\
NGC288   &  -0.045 & 0.002 & 10.097 & 9.32 &   1.16 & 133.8 & 12.0 & -1.32 \\
NGC362   &  -0.040 & 0.001 & 10.061 & 8.93 &   3.45 & 49.2  & 9.4  & -1.26 \\
NGC1261   &   0.023 & 0.001   & 10.061 & 9.12 &   1.67 & 40.8 & 18.1  & -1.27  \\
NGC1851   &   -0.032 &  0.001 & 10.079 & 8.82 &   3.02 & 30.6 & 16.6  & -1.18   \\
NGC1978 &  -0.081 & 0.003     & 9.301 & 9.02 &   2.00 & 31.1 & 49.6  & -0.35 \\
NGC2808  &  -0.029 & 0.001  & 10.079 & 9.15 &   7.42 & 49.0 & 11.1  & -1.14 \\
NGC5272   &  -0.059 & 0.001 & 10.097 & 9.79      &   3.94 & 186.0 & 12.0  & -1.5 \\
NGC5286  &  -0.013 & 0.001  & 10.114 &  9.11 &   4.01 & 43.8 & 8.9 & -1.69    \\
NGC6093    &   0.056 & 0.001 & 10.130 & 8.80 &   2.49 & 36.6 & 3.8  & -1.75    \\
NGC6101   &    -0.003 &  0.001    &  10.114  &  9.22 & 1.27 & 63.0  & 11.2  & -1.98 \\
NGC6362   &  -0.010 & 0.002 & 10.097  & 9.20 &  1.47 & 123.0 & 5.1 & -0.99 \\
NGC6584   &   0.033 & 0.002 & 10.088 & 9.02 &   0.91 & 43.8 & 7.0  & -1.50  \\
NGC6624   &   0.016 & 0.002  & 10.114 & 8.71 &  0.73 & 49.2 &  1.2 &  -0.44 \\
NGC6637   &  -0.028 & 0.001  & 10.097 & 8.82 & 2.45 &  50.4  & 1.7  & -0.64  \\
NGC6652   &   0.029 & 0.003   & 10.122 & 8.39  &  0.57 & 28.8 & 2.7 & -0.81  \\
NGC6681   &  -0.031 & 0.003  & 10.114 & 8.65 & 1.13 & 42.6  & 2.2  & -1.62 \\
NGC6715   &  -0.107 & 0.001  & 10.079  & 9.93 &  14.1 & 49.2  & 18.9  & -1.49 \\
NGC6717    &   0.080 & 0.004 &  10.114 & 8.22  & 0.36 & 45.0  & 2.4 & -1.26 \\
NGC6934  &   0.000 & 0.002  & 10.079 & 9.04 & 1.17 & 41.4  & 12.8 & -1.47\\
\enddata
\tablecomments{Ages are from \citet{dotter10} and \citet{martocchia18a,dalessandro16} for NGC1978 and NGC121.
Masses for Galactic GCs are from \citet{baumgardt18}, for NGC121 we used values from \citet{glatt11} and from \citet{krause16} for NGC1978. 
Relaxation times comes from \citet{harris96}, \citet{glatt11} for NGC121 and the present work for NGC1978.}
\end{deluxetable}

The authors thank the anonymous referee for the careful reading of the paper and the useful comments that improved the presentation of this work
E.D. acknowledges support from The Leverhulme Trust Visiting Professorship Programme VP2- 2017-030.
E.D. warmly thank Holger Baumgardt for providing the $M_{PD}/M_{ini}$ values.
E.D. also thank Francesco Calura and Michele Bellazzini for useful discussions.
The research is funded by the project Light-on-Dark granted by MIUR through PRIN2017-000000 contract (PI: Ferraro).
N.B. gratefully acknowledges financial support from the Royal Society (University Research Fellowship) and the European Research Council (ERC-CoG-646928, Multi-Pop).



\end{document}